\documentclass[referee]{raa}
\usepackage{graphicx,times}
\usepackage{natbib}
\usepackage{amssymb,amsmath}
\bibpunct{(}{)}{;}{a}{}{,}

\usepackage[letterpaper=true,pagebackref=true]{hyperref}
\hypersetup{pdftitle = The title of my PDF, pdfauthor = My name, pdfsubject= The subject, pdfkeywords = keyword1 keyword2 keyword3}
\hypersetup{colorlinks = true, linkcolor = green, anchorcolor = red, citecolor = blue, filecolor = red, pagecolor = red, urlcolor = red}

\begin{document}

 \title{Irreversible Rapid Changes of Magnetic Field Associated with the 2012 October 23 Circular Near-limb X1.8 Flare
}

   \volnopage{Vol.0 (200x) No.0, 000--000}         \setcounter{page}{1}          
   \author{Dandan Ye, Chang Liu
   \and Haimin Wang
   }

   \institute{Space Weather Research Laboratory and Big Bear Solar Observatory, New Jersey Institute of Technology, University Heights, Newark, NJ 07102-1982, USA; {\it haimin.wang@njit.edu}\\
   }

   \date{Received~~2016 month day; accepted~~2016~~month day}

\abstract{It has been found that photospheric magnetic fields can change in accordance with the three-dimensional magnetic field restructuring following solar eruptions. Previous studies mainly use vector magnetic field data taken for events near the disk center. In this paper, we analyze the magnetic field evolution associated with the 2012 October 23 X1.8 flare in NOAA AR 11598 that is close to the solar limb, using both the 45 s cadence line-of-sight and 12 minute cadence vector magnetograms from the Helioseismic and Magnetic Imager on board Solar Dynamic Observatory. This flare is classified as a circular-ribbon flare with spine-fan type magnetic topology containing a null point. In the line-of-sight magnetograms, there are two apparent polarity inversion lines (PIL). The PIL closer to the limb is affected more by the projection effect. Between these two PILs there lie positive polarity magnetic fields, which are surrounded by negative polarity fields outside the PILs. We find that after the flare, both the apparent limb-ward and disk-ward negative fluxes decrease, while the positive flux in-between increases. We also find that the horizontal magnetic fields have a significant increase along the disk-ward PIL,  while in surrounding area, they decrease. Synthesizing the observed field changes, we conclude that the magnetic fields collapse toward the surface  above the disk-ward PIL as depicted in the coronal implosion scenario, while the peripheral field turns to a more vertical configuration after the flare. We also suggest that this event is an asymmetric circular-ribbon flare: a flux rope is likely present above the disk-ward PIL.  Its eruption causes the instability of the entire fan-spine structure and the implosion near that PIL.}


   \authorrunning{ }               \titlerunning{Magnetic Field Changes Associated with a Circular Ribbon Flare}     \maketitle

\section{Introduction}           \label{sect:intro}

The long-term photospheric magnetic field evolution can play an important role in solar eruptions like flares, coronal mass ejections (CMEs), and filament eruptions as shown in many previous studies (see the review by \citet{WL15} and \citet{Yang15}). Because of much higher gas pressure and density in the photosphere comparing to the corona, photospheric magnetic fields do not change significantly during eruptions as usually found in earlier observations. This has been considered as one of the basic assumptions on modeling of eruptions.

Using ground-based observations, \citet{W92} and \citet{W94} first found rapid changes of vector magnetic fields associated with some X-class flares. Later, many studies have confirmed the earlier results and advanced our knowledge of flare-related photospheric field changes (see the review by \citet{WL15}).
Most notably, \citet{WS12} proved that photospheric magnetic fields near the polarity inversion line (PIL) became more horizontal immediately following eruptions. The authors also suggested that the line-of-sight (LOS) magnetograms also provide a useful tool for understanding the topological changes of active region magnetic fields associated with eruptions. Indeed, studies have shown that the disk-ward flux decreases while the limb-ward flux increases right after flares \citep[e.g.,][]{W02, W06, WL10}. \citet{Sudol05} also found that photospheric magnetic fields changed permanently during eruptions by using GONG magnetograms. They employed a step function to characterize the rapid changes of magnetic field. \citet{wangj09} studied the X7.1 flare close to the solar limb,  and found clear evidence of weakening in the horizontal magnetic fields in peripheral area of the $\delta$-sunspot group and strengthening of it  in an extended area centralized at the PIL between major sunspots of opposite polarities. This demonstrated the usefulness of the projection effect of near-limb events-- it avoids the complicated analysis of vector magnetograms in deriving the horizontal components of vector magnetic fields.

There has been recent progress in the study of magnetic field changes associated with flares \citep{wangh14,WL12,P14}. Importantly, such a phenomenon also motivated some modeling efforts. \citet{Hudson08} presented a model to predict that the photospheric magnetic field would become more horizontal after flares.
In addition, \citet{Hudson08} and \citet{Fisher12} used the concept of momentum conservation and assessed the Lorentz-force changes due to the release of energy associated with the rapid corona magnetic field evolution. It is argued that a back-reaction following CME eruptions could push the photospheric magnetic fields to become more horizontal near the flaring PIL. The authors estimated the change of the integrated vertical Lorentz force exerted on the photosphere from the corona to be

\begin{equation}
\delta F_r = \frac{1}{8\pi}\int(\delta B_r^2 - \delta B_h^2)dA,
\end{equation}
where $B_r$ is the vertical field strength, $B_h$ is the horizontal field strength, and $dA$ is the surface of vector magnetogram. The integration area is denoted by the region of interest (ROI) as we discuss below.

The above observational studies and theoretic modeling are mainly based on the classical two-ribbon flare topology. Recently, it become evident that many flaring regions have a spine-fan magnetic structure and produce so-called circular-ribbon flares \citep[e.g.,][]{WL12,wangh14,liu15,joshi15,WL15}. Understanding this type of events requires extending the magnetic topology analysis from 2D to 3D. For the X1.8 flare on 2012 October 23 under the present study, \citet{Yang15} found that it has many characteristics of circular-ribbon flares. The main topological structures of the flaring region as revealed by the authors include a null point embedded in the fan-spine fields, a flux rope lying below the fan dome, and a large-scale quasi-separatrix layer induced by the quadrupolar-like magnetic field of the active region. These motivated us to study the rapid 3D magnetic field changes associated with this particular circular-ribbon flare.

Although many previous studies have provided evidence of the relationship between magnetic field changes and solar eruptions, they usually focus on just one component of photospheric magnetic field data. Thus a complete understanding of the field restructuring is still not achieved. In this study, to examine magnetic evolution associated with the circular-ribbon flare, we will study both the LOS and vector photospheric magnetic field observations. In Section 2, we introduce the observation and data reduction procedure. In Section 3, we present the results for magnetic field changes. The explanation about the flare-associated magnetic field evolution and restructuring are given in Section 4.

\section{Observation and Data Reduction}

The X1.8 flare on 2012 October 23 occurred in NOAA active region 11598, located close to the east limb (S10$^{\circ}$E42$^{\circ}$). The flare started at 03:13 UT, peaked at 03:17 UT, and ended at 03:21 UT in \textit{GOES} 1--8~\AA\ soft X-ray flux. Full-disk, multi-wavelength observation with high resolution have been made available with the Helioseismic and Magnetic Imager (HMI, \citep{Schou12}) and the Atmospheric Imaging Assembly (AIA) on board the Solar Dynamics Observatory (SDO). The LOS magnetic field observation from HMI has a pixel of $\sim$0$\farcs$5 and a cadence of 45 s. The noise level of the LOS field measurement is about 10 G. We also analyzed HMI vector magnetic field data (\citet{Hoek14}), which have the same pixel scale but a lower cadence (12 minutes) compared to the LOS observation. As we mentioned earlier, \citet{Yang15} focused on the same active region in their study. Mainly based on magnetic topology and flare emission morphology as seen in AIA images and magnetic field models, the authors suggested that this event is a circular-ribbon flare.

Before quantitatively analyzing the data, we first scrutinized movies of HMI intensity, LOS and vector magnetic fields, and AIA EUV images covering the flare. Similar to the event analyzed by \citet{liu15}, the circular ribbon in this flare is somehow asymmetric: under part of the dome structure closer to disk center, a filament flux rope is embedded. This is also the core part of the $\delta$ configuration of the flaring region under study. \citet{Yang15} presented the flare ribbons, the dome structure of magnetic fields and magnetic flux rope nicely for this event (see Figures 2, 3 and 5 in their paper). From the HMI intensity movie, we can observe a rapid enhancement (darkening) of sunspot feature at the central location of this $\delta$ structure, together with the decay of penumbrae in the surrounding areas. To pinpoint the location of magnetic field changes, we constructed difference images for both LOS and horizontal field magnetograms, which will be discussed in the next section.
From the difference images, we can easily identify ROIs for our analysis, which covers the region with most obvious magnetic field changes.  Our study will focus on the time variation of different parameters at different ROIs, in order to evaluate the rapid/irreversible changes of magnetic fields associated with the flare.

\section{Results}
\subsection{The change of LOS magnetic fields}

In Figure 1, we demonstrate the overall structure of this active region and its flare-related changes. The top panels show the LOS magnetograms before and after the flare and their difference image (the postflare image minus the preflare image). The bottom panels show the corresponding intensity images. The red boxes outline the main ROIs, which include three magnetic components: Area I that contains negative flux in the left  box (limb-ward negative flux),  Area II that contains all positive flux in both red boxes; Area III that contains all the negative flux in the right box (disk-ward negative flux). Two apparent PILs are observed, and we name them the disk-ward PIL and the limb-ward PIL  (labeled as PIL 1 and PIL 2 later in Figure 6). As \citet{wangh14} pointed out,  this kind of structure of double PILs may be closely associated with quasi-circular flares. A filament flux rope is found to be present in this flaring region, lying above the disk-ward PIL \citep{Yang15}. The difference images (the right column) show that changes of the apparent magnetic field and intensity are concentrated in the marked ROIs.

Figure 2 plots different components of the LOS magnetic flux as a function of time. The top red curve shows the evolution of the central positive flux (Area II). The absolution value of the disk-ward negative flux, which is in Area III, is shown as the blue curve in the middle panel and the limb-ward negative flux in Area I is in the bottom. It is obvious that closely associated with the occurrence of the flare that peaked around 03:17~UT, all these three magnetic flux components exhibit significant changes in a step-like fashion. Specifically, the flux in the central positive polarity region increases by about 45\% after the flare; meanwhile, negative fluxes decrease by about 26\% and 20\% in the disk-ward and limb-ward negative flux regions, respectively.

\begin{figure}
   \centering
  \includegraphics[width=15.5cm, angle=0]{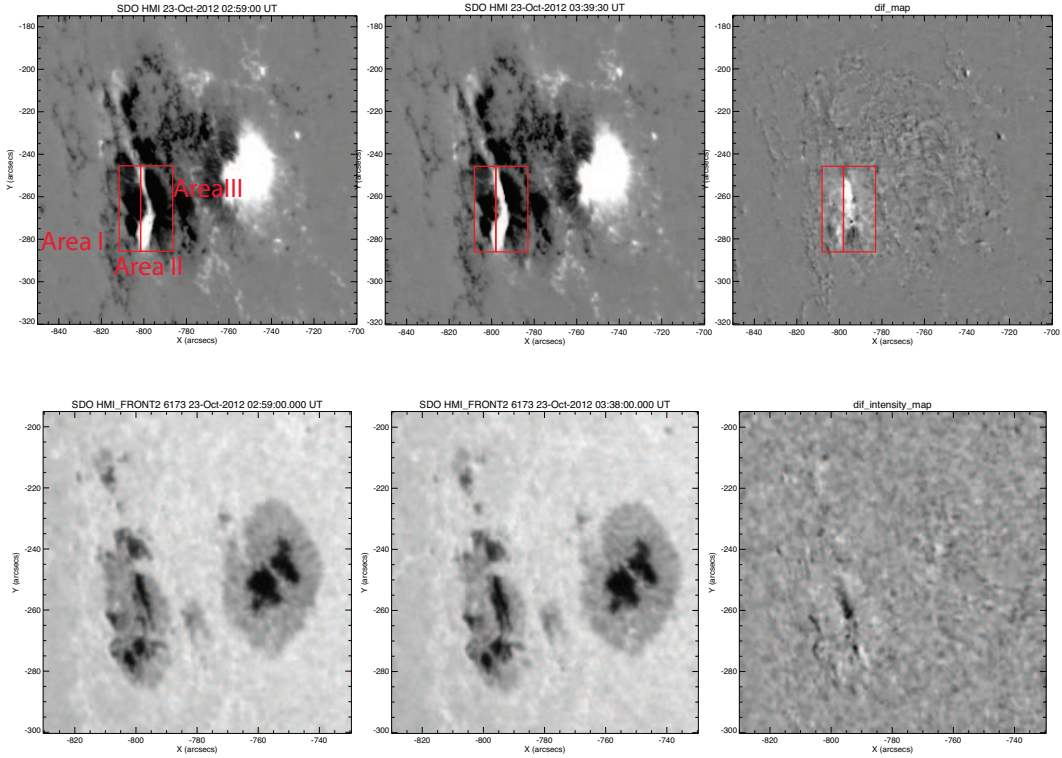}
     \caption{\emph{Top panels}: HMI magnetograms before and after the X1.8 flare on 2012 October 23, and their difference image.  \emph{Bottom panels}: HMI intensity images before and after the flare, and their difference image. The red boxes outline the region of the circular-ribbon flare, featuring double PILs. Three components magnetic flux are defined. The negative flux in left box is defined as Area I, the positive flux in both boxes as Area II, and negative flux in right box as Area III.}
   \label{Fig1}
   \end{figure}

\begin{figure}
   \centering
  \includegraphics[width=14.5cm, angle=0]{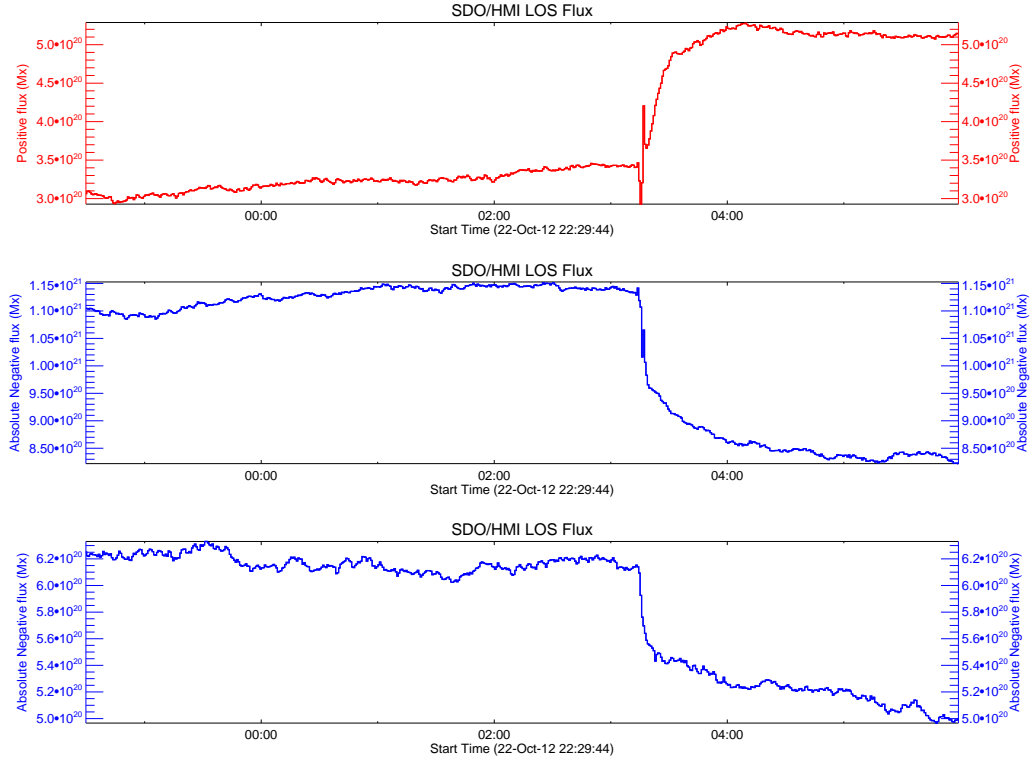}
     \caption{The SDO/HMI LOS flux as a function of time. \emph{Top}: The central positive flux (Area II). \emph{Middle}: The disk-ward negative flux in the absolute value (Area III). \emph{Bottom}: The limb-ward negative flux in the absolute value (Area I).}
   \label{Fig2}
   \end{figure}

\subsection{The change of vector magnetic fields}

For vector field analysis, we use vector magnetograms prepared by the HMI team. The 180$^\circ$ azimuthal ambiguity is resolved and the observed fields are also transformed to heliographic coordinates so that the projection effect is removed. We understand the difficulty and uncertainty of these data analysis procedures.  That is why  we also used directly observed LOS fields to provide complementary information to understand 3-D magnetic field evolution for this near-limb event. As shown in Figure 3, the surface magnetic field structure typical for circular-ribbon flares is more obvious: the central parasite positive flux is surrounded by negative flux. Here we focus on the study of evolution of the horizontal field component parallel to the solar surface, which can be calculated as $B_h=\sqrt{B_x^2+B_y^2}$.

Similar to the analysis of LOS field and intensity, we subtract the horizontal magnetogram after the flare by the horizontal magnetogram before the flare to obtain the difference horizontal field image (bottom right panel of Figure 3). We can see that the most obvious change is the enhancement of horizontal field at the location of the disk-ward PIL (right edge of the green box) where the flux rope is embedded. Also, this enhancement is surrounded by regions of decreased horizontal field. The temporal evolutions of the increased and decreased mean field strength are shown in Figures 4 and 5. The results show that after the flare, the mean horizontal field in the disk-ward PIL region increases by about 190~G, i.e., about 25\% of the pre-flare value. The mean horizontal field in the surrounding region decreases by about 130~G, i.e. about 13\% of the pre-flare value. Using Equation (1), we derive that associated with the flare, the total Lorentz force change in the vertical direction is about $3.9 \times 10^{22}$~dyne, comparable with previous studies \citep{Hudson08,Fisher12,Shuo12}. The maximum change of Lorentz force per area is $3.4 \times 10^4$~dyne~cm$^{-2}$ and the mean value is $2.8 \times 10^4$ dyne~cm$^{-2}$.

\begin{figure}
   \centering
  \includegraphics[width=14.5cm, angle=0]{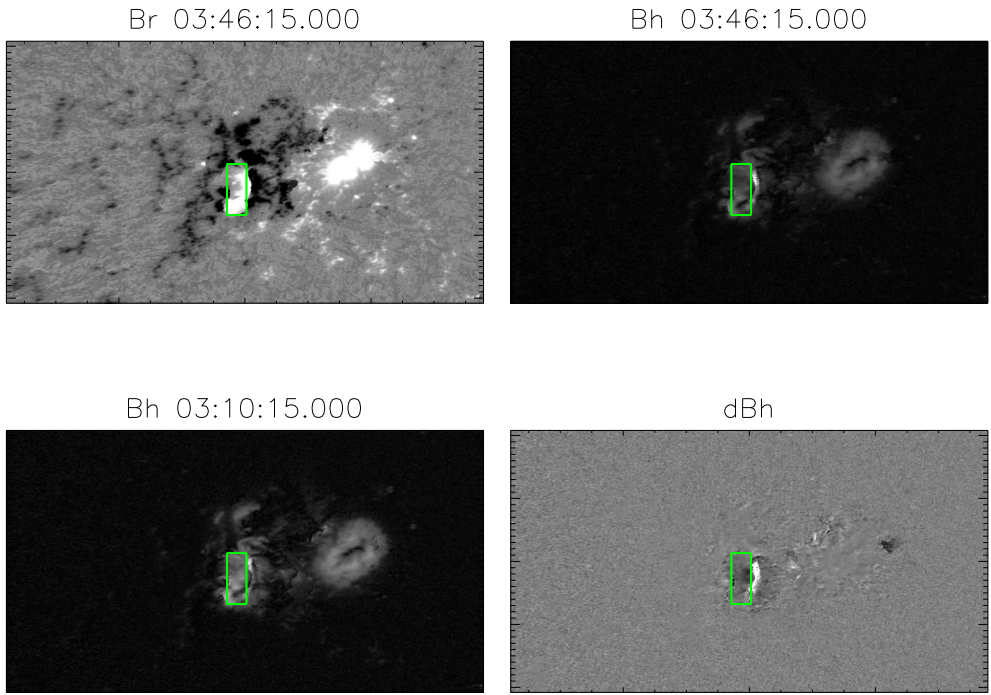}
     \caption{\emph{Top left}: Vertical component of the deprojected HMI vector magnetic field. \emph{Top right}: Horizontal component of vector magnetic field after the flare. \emph{Bottom left}: Horizontal component of vector magnetic field before the flare. \emph{Bottom right}: Difference image of horizontal magnetic field (the postflare image minus the preflare image). The green box outlines the center positive flux region.  The disk-ward PIL is next to the right edge of the box, where the horizontal field has obvious increase after the flare.}
   \label{Fig3}
   \end{figure}

\begin{figure}
   \centering
  \includegraphics[width=14.5cm, angle=0]{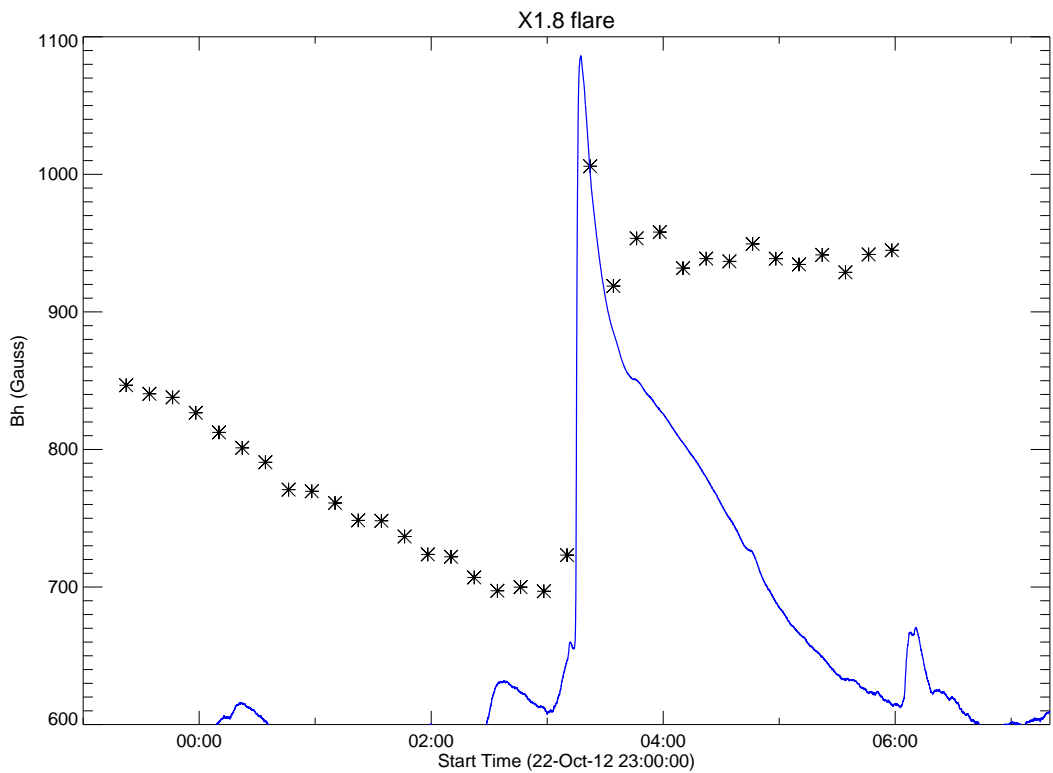}
     \caption{Time profile of horizontal magnetic field at the disk-ward PIL (center white area in Figure 3).  The blue curve shows the GOES 1--8~\AA\ soft X-ray flux, indicating the time of the flare.}
   \label{Fig4}
   \end{figure}

   \begin{figure}
   \centering
  \includegraphics[width=14.5cm, angle=0]{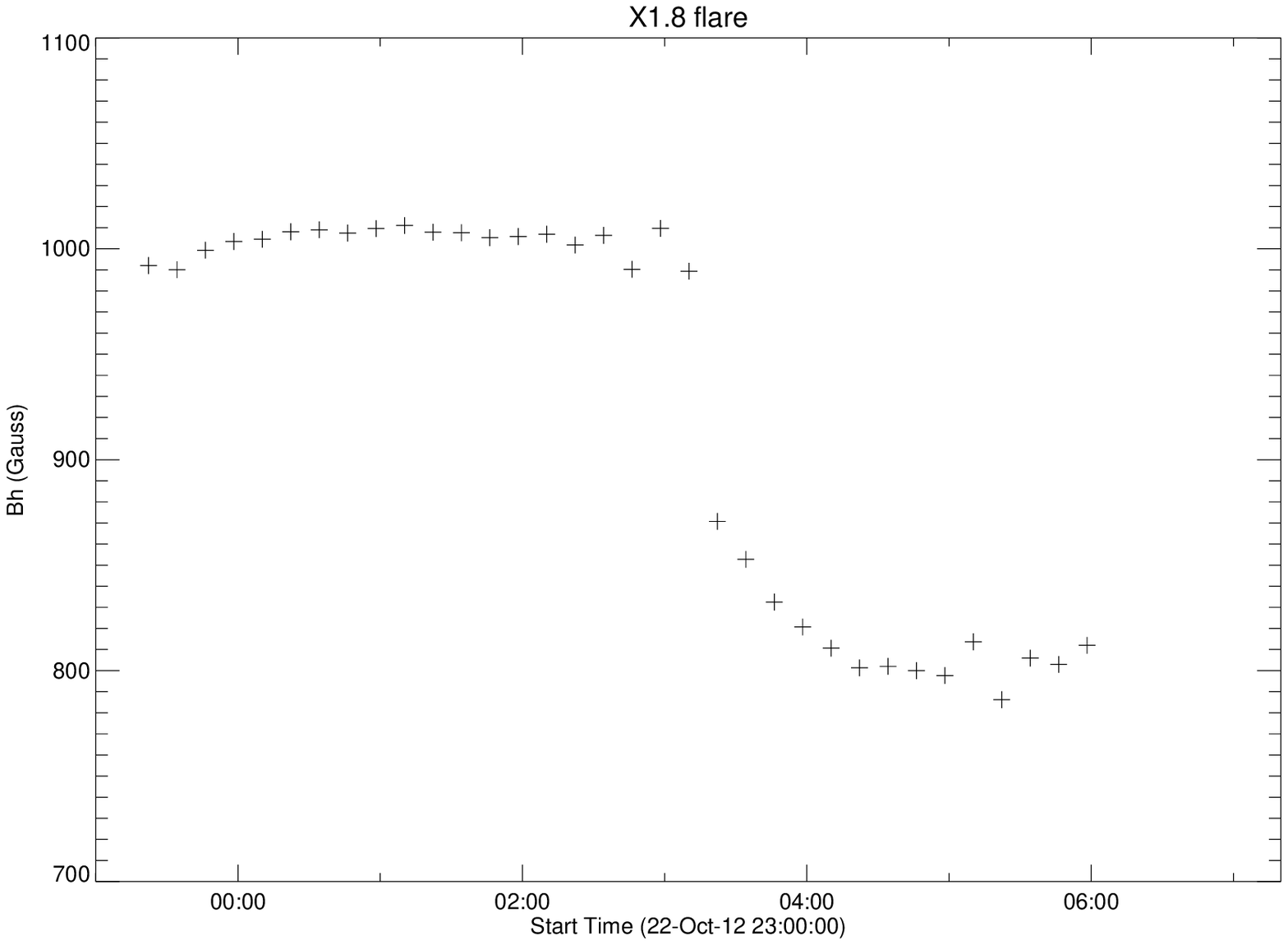}
     \caption{Time profile of horizontal magnetic field in the area surrounding the disk-ward PIL (surrounding darkened area in Figure 3).}
   \label{Fig5}
   \end{figure}

\section{Summary and Discussion}
We investigated the flare-associated rapid changes of the LOS and horizontal components of magnetic fields associated with the X1.8 flare on 2012 October 23. This is a unique event as it is a circular-ribbon flare located close to the limb. The main results are summarized as follows.

1. For the LOS component, we found a rapid increase in the central positive flux and decrease in the surrounding negative fluxes (both disk-ward and limb-ward) after the flare.

2. For the horizontal component, we found a significant enhancement of magnetic field at the disk-ward PIL. This is also the location of an embedded flux rope as suggested by \citet{Yang15}.

All these results support the theory and prediction as elaborated by \citet{Hudson08}. As discussed in \citet{WL10}, the surrounding magnetic field lines would change to a more vertical state after eruptions, because the pressure in the region of flux rope eruption is released after the flare. In the meantime, the magnetic field lines at the flaring PIL would change to be more horizontal relative to the solar surface due to the back-reaction of the upward eruption. In Figure~6, we use a simple cartoon picture to explain our observations. Since this is a circular-ribbon flare, there are two loop systems in the cross section of the fan-dome as illustrated in this 2D view. As we described above, the most significant changes are associated with the eruption of the flux rope located at the disk-ward PIL (PIL1). The limb-ward system above PIL2 expands outward in response to the flux rope eruption, which can explain why the limb-ward negative flux also decreases. In summary, we believe that this event is similar to the flare studied by \citet{liu15}: the flux rope eruption destabilizes the fan-spine structure and triggers the null point reconnection. The enhancement of horizontal field at PIL1 is only a consequence of the flux rope eruption, not the null point reconnection.

\begin{acknowledgements}

We thank the referee for providing valuable comments to improve the paper.  We acknowledge the SDO/HMI team for the magnetic field data. This work was supported by US NSF under grants AGS 1348513 and 1408703.

\end{acknowledgements}

\begin{figure}
   \centering
  \includegraphics[width=14.5cm, angle=0]{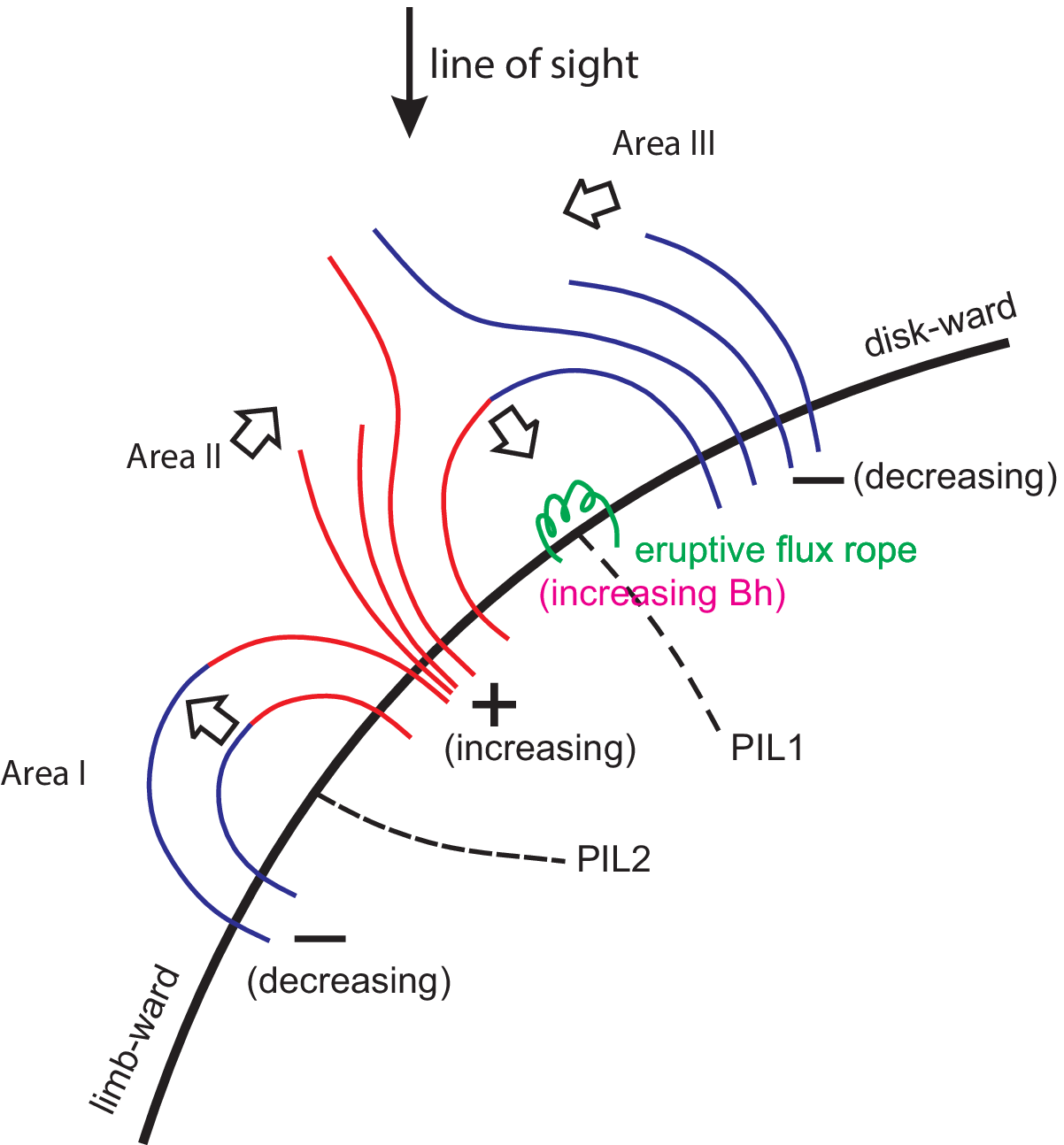}
     \caption{A cartoon picture interpreting the observed magnetic field changes associated with the 2012 October 23 X1.8 circular-ribbon flare. The eruption of a flux rope above PIL1 causes coronal implosion and the subsequent enhancement of the horizontal field at PIL1. The surrounding fields collapse toward the center to increase the LOS field strength in the central positive flux region and also to decrease field in the disk-ward negative flux region. The loops above PIL2 also expand outward, explaining the decrease of the limb-ward negative flux, which also contributes to the central positive flux increase.}
   \label{Fig6}
   \end{figure}

\end{document}